\documentclass[seceq]{jpsj2} 
\title{
Correlation Structures of Correlated Binomial Models and
Implied Default Distribution}

\author{
Shintaro \textsc{Mori}$^{1}$
\thanks{E-mail: mori@sci.kitasato-u.ac.jp},
Kenji \textsc{Kitsukawa}$^{2}$
\thanks{E-mail: kenji.kitsukawa@daiwasmbc.co.jp},
 and Masato \textsc{Hisakado}$^{3}$\thanks{E-mail address:
hisakado\_masato@standardandpoors.com}}

\inst{$^{1}$
Department of Physics, School of Science,
Kitasato University,\\
 Kitasato, Sagamihara,
Kanagawa 228-8555 
\\
$^{2}$
Daiwa Securities SMBC,  
Marunouchi 1-9-1, Chiyoda-ku, Tokyo 100-6753
\\
$^{3}$Standard \& Poor's,
Marunouchi 1-6-5, Chiyoda-ku, Tokyo
100-0005 \\
}

\abst{
We show how to analyze  and interpret the correlation structures, 
the conditional expectation values and correlation 
coefficients of exchangeable Bernoulli random variables.
We study implied default distributions 
for the iTraxx-CJ tranches and  some popular probabilistic models, 
including the Gaussian copula model,
Beta binomial distribution model and 
long-range Ising model. 
We interpret the differences in their 
profiles in terms of the correlation structures.
The implied default distribution has singular correlation structures,
 reflecting the credit market implications.
We point out two possible origins of the singular behavior. 
}

\kword{
correlation, calibration, Beta binomial, Gaussian copula model, default,
dependency structure, Ising model
}

\begin{document}
\maketitle

\section{Introduction}

Describing and understanding crises in markets 
are intriguing subjects in financial engineering 
and econophysics 
\cite{Schonbucher, Bouchaud, Stanley,Davis,Kitsukawa,Sakata}. 
In the context of econophysics, the mechanism of systemic failure in 
banking has been studied \cite{Aleksiejuk,Iori}.
The Power law distribution of 
 avalanches and several scaling laws in the context of the percolation theory
were found.
In addition, the network structures of real companies have been studied   
recently and their nonhomogeneity nature have 
been clarified \cite{Souma, Chmiel,Helbing}. 
This feature  should be taken into account in the modeling 
of the dependent defaults of companies.  

In financial engineering, many products have been
invented to hedge the credit risks. CDS is a single-name credit derivative
 which is targeted on the default of one single obligor.
Collateralized debt obligations (CDOs) are
financial innovations to securitize portfolios of defaultable assets,
which are called credit portfolios.
They provide protections against a subset  of total loss on a credit 
portfolio in exchange for payments.
From an econophysical viewpoint, they give valuable insights 
into the market implications on default dependencies and the clustering of 
defaults. 
This aspect is very important, because the main difficulty  in 
the understanding of credit events is that we have no sufficient
information about them. By empirical 
studies of  the historical data on
credit events, 
 the default probability $p_{d}$ and default correlation $\rho_{d}$ 
were estimated \cite{Jobst}.
 However, more detailed information is necessary in the pricing of
 credit derivatives and in the evaluation of models in
 econophysics. 
 The quotes of the CDOs depend on the profiles of the default
 probability function \cite{Finger}. This means that 
it is possible to infer the default loss probability function 
from market quotes. 
Recently, such an ``implied'' loss distribution function has  
attracted much attention in the studies of  credit derivatives.
 Instead of using popular probabilistic  models,
 implied loss distribution are proposed to 
use \cite{Vacca,Hull4}. 

In this paper, we show how to get detailed information
contained in probability functions for multiple defaults. 
We compare the implied loss probability function with some popular 
probabilistic models and show their differences in terms of the 
correlation structure.  
The paper is organized as follows.
In \S  \ref{correlation} we start from the definition
of exchangeable Bernoulli random variables and 
explain the term   
``correlation structures'', the conditional expectation values and 
correlations. We introduce several notations of related quantities.  
Using the recursive relations, we show how to  estimate them.
We also point out that the  method can be applied to any probability 
function of Bernoulli random variables, 
that are not necessarily exchangeable.
In \S \ref{entropy}, 
we show how to infer the loss probability 
function for multiple defaults from the CDO market quotes by the 
entropy maximum principle.
We compare the 
implied loss probability function with those of some popular 
probabilistic models
in \S \ref{comparison}. The differences 
become strongly apparent in the behavior of the conditional
correlations. 
The singular behavior of the implied loss function should be
attributed to the nonlinear nature of multiple defaults or 
network structures of companies.
We also try to read the credit market 
implications contained in the market quotes of CDOs and 
make a comment on
``Correlation Smile''.
Section \ref{conclusion} is devoted to the  
summary and future problems.
In the appendix, 
we explain the relation between the profiles of  
probability functions and the correlation structures.

\section{Calibration of Correlation Structures}
\label{correlation}

In this section, we show a method of 
obtaining the ``correlation structure'' from 
the probability function.
 We denote the i-th asset's (or obligor's) state by Bernoulli random 
variable $X_{i}=0,1\hspace*{0.2cm}(i=1,\cdots,N)$. 
If the asset is
defaulted (or non-defaulted), $X_{i}$ takes $1 (\mbox{resp.} 0)$.
We assume
that $X_{i}$s are exchangeable. 
The exchangeability means that 
the joint probability 
function of $X_{i}$s is independent of  any permutation of the values of 
$X_{i}$s. Denoting the joint probability function as 
\[
\mbox{Prob.}(X_{1}=x_{1},X_{2}=x_{2},\cdots,X_{N}=x_{N})=P(x_{1},x_{2},\cdots,x_{N}) ,
\]
the next relation holds for any permutation 
$i_{1},i_{2},\cdots,i_{N}$ of $1,2,\cdots,N$,
\[
P(x_{1},x_{2},\cdots,x_{N})
=P(x_{i_{1}},x_{i_{2}},\cdots,x_{i_{N}}).
\]
By assumption, the remaining degree of freedom in the joint 
probability function reduces to $N$. 
The joint probability for $i$ defaults 
and $j$ nondefaults only depends only on $i$ and $j$,
 and we denote it as $X_{i,j}$. The probability function for $n$
 defaults $P_{N}(n)$ is written as 
\[
P_{N}(n)={}_{N}C_{n}\cdot X_{n,N-n}. 
\]
Here ${}_{N}C_{n}$ is the binomial coefficients.

The term ``correlation structure'' means the conditional 
expectation values $p_{i,j}$  and 
correlations $\rho_{i,j}$ . 
The subscript ${}_{i,j}$ of $p_{i,j}$ and 
$\rho_{i,j}$
means that
they are estimated under the condition that any $i$ (resp.$j$) of $N$ variables
tale 1 (resp. 0). 
We also introduce $q_{i,j}$ as $1-p_{i,j}$.
$p_{0,0}$ is the unconditional expectation value and it is nothing but
the default probability $p_{d}$. $\rho_{0,0}$ is the unconditional default 
correlation $\rho_{d}$. More detailed explanations about $p_{i,j}$ and
$\rho_{i,j}$ are given in the appendix.
These quantities satisfy the following relations \cite{Hisakado}
\begin{eqnarray}
p_{i+1,j}&=&p_{i,j}+(1-p_{i,j})\rho_{i,j}, \label{consist_p}\\
q_{i,j+1}&=&q_{i,j}+(1-q_{i,j})\rho_{i,j},   \label{consist_q} \\
p_{i-1j}-p_{i,j-1}&=&-(1-p_{i-1j})\rho_{i-1j}-p_{i,j-1}\rho_{i,j-1}
\label{consist_rho}. 
\end{eqnarray}
\begin{figure}[htbp]
\begin{center}
\includegraphics[width=7.0cm]{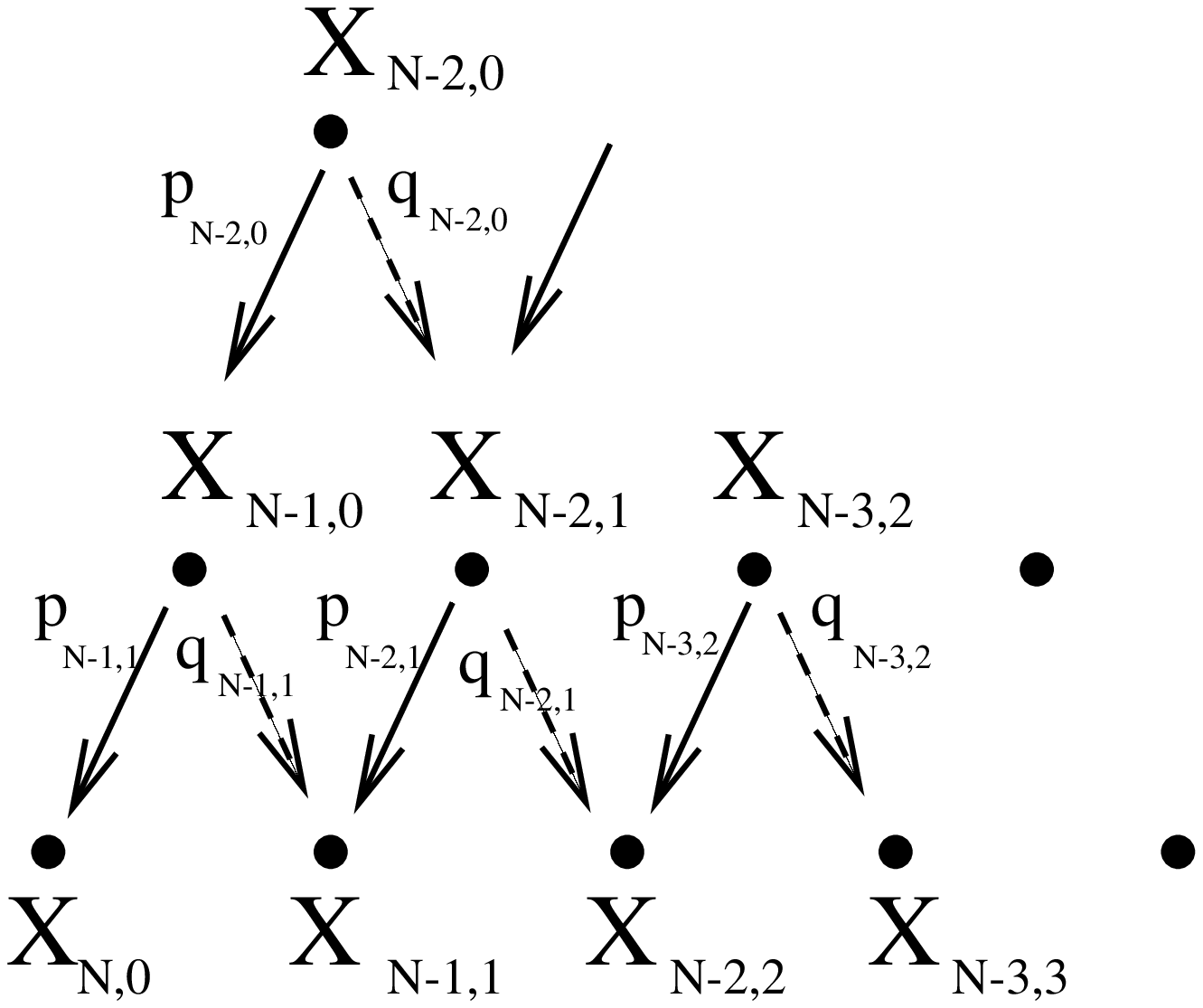}
\caption{\label{fig:Pascal_2}Solving process for $p_{i,j},q_{i,j}$ and
 $\rho_{i,j}$ from $X_{n.N-n}$.}
\end{center}
\end{figure}

Using these recursive relations, it is possible to 
 estimate $p_{i,j}$ and $\rho_{i,j}$ from $P_{N}(n)$ or $X_{n,N-n}$. 
$X_{n,N-n}$ are on the bottom line of the 
Pascal triangle (See Fig.\ref{fig:Pascal_2}). 
Then recursively solving the above eqs.
(\ref{consist_p})-(\ref{consist_rho})
to the top vertex $(0,0)$ of the Pascal triangle, 
we obtain all $p_{i,j}$s  and $\rho_{i,j}$s. 
For example, to obtain $p_{N-1,0}$ we use the
relations
\begin{equation}
X_{N,0}=X_{N-1,0}\cdot p_{N-1,0} \hspace*{0.3cm}\mbox{and}\hspace*{0.3cm}
X_{N-1,1}=X_{N-1,0}\cdot (1-p_{N-1,0}) .
\end{equation}
Solving for $p_{N-1,1}$, we get
\begin{equation}
p_{N-1,0}=\frac{X_{N,0}}{X_{N,0}+X_{N-1,1}}.
\end{equation}
Likewise, we can estimate $p_{i,j}$ for general $i,j\le N-1$.
From $p_{i,j}$, $\rho_{i,j}$ are obtained by solving eq. (\ref{consist_p}).

The important point is that it is possible to estimate 
the correlation structure for any $P_{N}(n)$. In addition to 
theoretical models, empirically obtained probability functions
can be studied. 
If $X_{i}$s are exchangeable, the obtained $p_{i,j}$ and $\rho_{i,j}$ 
are the ones defined in the text.  
In network terminology, the exchangeable case corresponds to the 
complete graph $K_{N}$ where all nodes are similar to each other 
with the same strength. 
If the network structure is not uniform,
 $X_{i}$s are not exchangeable. 
Even in such a case, the above method is applicable and 
gives many insights into the system. For example, if the 
network structure is treelike, 
the obtained correlation structure should be  completely 
different from that in the exchangeable case.
 Its singular behavior strongly suggests the
nonuniform network structure of the system.

\section{Implied Default Distribution}
\label{entropy}
 
We show how to infer the loss probability function 
based on market quotes of CDOs \cite{Vacca,Hull4}. 
In advance, we briefly explain CDOs.
CDOs provide protection against losses in credit portfolios.
Here ``credit'' means that the constituent assets of the portfolio 
can be defaulted. If an asset is defaulted, the portfolio 
loses its value.  
The interesting point of CDOs  is that they are divided into 
several parts (called 'tranches'). Tranches have 
priorities that are defined by the attachment point $a_{L}$
 and the detachment point $a_{H}$. The seller of protection agrees to cover
all losses between $a_{L}K_{Total}$ and $a_{H}K_{Total}$, where 
$K_{Total}$ is the initial total notional of the portfolio.
That is, if the loss is below $a_{L}K_{Total}$, the tranche does not
cover it. 
The tranche begins to cover it
Only when it exceeds $a_{L}K_{Total}$. 
If it exceeds $a_{H}K_{Total}$, the notional becomes zero.
The seller of protection receives payments at a rate $s$
on the initial notional $(a_{H}-a_{L})K_{Total}$. Each loss that is
covered reduces the notional on which payments are based.
A typical CDO 
has a life of 5 years during which the seller of
protection receives periodic payments. Usually these payments 
are made quarterly in arrears. In addition, to bring 
periodic payments up to date, an accrual payment is performed.
Furthermore, the seller of protection makes a payment equal to the 
loss to the buyer of protection. The loss is the
reduction in the notional principal times one less the 
recovery rate $R$.

iTraxx-CJ is an equally weighted portfolio of fifty CDSs on Japanese
 companies. The notional principal of CDSs is $K$ and $K_{Total}$ is
$50\hspace*{0.1cm} K$. The recovery rate is $R=0.35$.
The standard attachment and detachment points are
$\{0\%,3\%\}$,$\{3\%,6\%\}$,$\{6\%,9\%\}$,$\{9\%,12\%\}$ and
 $\{12\%,22\%\}$.
We denote them as $\{a_{L}^{i},a_{H}^{i}\}$ with $i=1,\cdots,5$.
Table \ref{tab:quotes} shows the tranche structures and 
quotes for iTraxx-CJ (Series 2) on August 30, 2005.
We denote the upfront payment as $U_{i}$ and the annual payment rate
as $s_{i}$ in basis points per year for the $i$th tranche. 
In the last row, we show the data for the index that cover all
 losses for the portfolio. In the 6th column, we show the initial
 notional $N_{0}^{i}$ in units of $K$.

\begin{table}[htbp]
\caption{\label{tab:quotes} Quotes for iTraxx-CJ (Series 2) on 
August 30, 2005. Quotes are in basis points.
Source: Tranche, Morgan Stanley Japan Securities Co. and Index, Bloomberg
}
\begin{center}
\begin{tabular}{ccccccc}
Tranche  & $a_{L}^{i}$ & $a_{H}^{i}$ & $s_{i}$[bps] & $U_{i}$[bps] &
 $N_{0}^{i}$& $N_{T,Implied}^{i}$ \\
\hline
1 & 0\% & 3\% & 300 & 1313.3 & 1.5 & 1.1066 \\
2 & 3\% & 6\% & 89.167 & 0 & 1.5   & 1.4361 \\
3 & 6\% & 9\% & 28.5 & 0 & 1.5     & 1.4792 \\
4 & 9\% & 12\% & 20.0 & 0 & 1.5    & 1.4854 \\
5 & 12\% & 22\% & 14.0 & 0  &  5.0 & 4.9660 \\
6 & 0\% & 100\% & 22.08 & 0 &  50  & 49.464 \\
\end{tabular}
\end{center}
\end{table}

The value of contract is the present value of the expected cash flows.
For simplicity, we treat 5 years as one term and write $T=5 [year]$.
We also assume that defaults occur in the middle of the period.
We denote 
the notional principal for the $i$th tranche 
outstanding at maturity as $N_{T}^{i}$. 
The expected payoff of contract is 
\begin{equation}
U_{i}N_{0}^{i}+T <N_{T}^{i}> s_{i}
 e^{-rT}+(N_{0}^{i}-<N_{T}^{i}>)\frac{s_{i}T}{2}e^{-r\frac{T}{2}}
\label{A+B}.
\end{equation}
Here, $<A>$ means the expectation value of $A$  
and $r$ is the risk-free
rate of interest.
The expected loss due to default is 
\begin{equation}
(N_{0}^{i}-<N_{T}^{i}>)e^{-r\frac{T}{2}} \label{C}.
\end{equation}
The total value of the contract to the seller of protection
 is given by eqs. (\ref{A+B})-(\ref{C}).
Risk neutral values of $s_{i}$ and $U_{i}$ are determined so that
 eq. (\ref{A+B}) equals eq. (\ref{C}). 
Conversely, the market quotes for $s_{i}$ and $U_{i}$ 
tell us about the expected notional principal $<N_{T}^{i}>$.
We write them as $N_{T,Implied}^{i}$. 
The last column in Table \ref{tab:quotes} shows them  from the
market quotes $s_{i}$ and $U_{i}$.

$N_{T}^{i}$ are random 
variables and are related to the number of
defaults $n$ at maturity as
\begin{equation}
N_{T}^{i}(n)=
\left\{
\begin{array}{cc}
N_{0}^{i} & n< \lceil \frac{a_{L}^{i}N}{1-R} \rceil \\
a_{H}N-n(1-R)  & \lceil \frac{a_{L}^{i}N}{1-R} \rceil \le n < \lceil
 \frac{a_{H}^{i}N}{1-R} \rceil  \\
0  &  n \ge \lceil \frac{a_{H}^{i}N}{1-R} \rceil . \\
\end{array}
\right.
\end{equation}
Here, $\lceil x \rceil$ means the smallest integer greater than $x$.
To calculate the expectation value of $N_{T}^{i}(n)$, the default
probability function $P_{N}(n)$ is necessary. Inversely, using the
data on these expectation values $N_{T,Implied}^{i}$, we try to 
infer 
$P_{N}(n)$ from the maximum entropy principle.  
It states that one should consider the model $P_{N}(n)$ that maximizes
 the entropy functional subject to the conditions imposed by
previous known information.

The entropy 
functional $S[P_{N}(n)]$ is defined as
\begin{eqnarray}
&&S[P_{N}(n)]=\sum_{n=0}^{N}{}_{N}C_{n}\cdot  X_{n,N-n}\log
 X_{n,N-n} \nonumber \\
&+&\sum_{i=1}^{6}\lambda^{i}(\sum_{n=0}^{N}{}_{N}C_{n} \cdot X_{n,N-n}
N_{T}^{i}(n)-N_{T, Implied}^{i}). \label{eq:entropy}
\end{eqnarray}
In order to impose the condition 
$<N_{T}^{i}>=N_{T,Implied}^{i}$ on $P_{N}(n)$,
we introduce six Lagrange multipliers $\lambda^{i}$.
By maximizing \ref{eq:entropy}, we get the implied joint 
probability $X_{n,N-n}$ as
\begin{eqnarray}
&&X_{n,N-n} \propto  
\left\{
\begin{array}{cc}
e^{-\lambda_{6}n-\lambda^{1}(a_{H}^{1}N-n(1-R))}\prod_{i=2}^{5}C_{i}
 & n<  n_{H}^{1} \\
e^{-\lambda_{6}n-\lambda^{j+1}(a_{H}^{j+1}N-n(1-R))}
\prod_{i=j+2}^{5}C_{i}
 & n_{H}^{j} \le n<  n_{H}^{j+1}   \\ \label{eq:Dist}
e^{-\lambda_{6}n} &  n \ge n_{H}^{5}  . \\   
\end{array}
\right.
\end{eqnarray}
Here, we use the notation $n_{H}^{i}=\lceil \frac{a_{H}^{i}N}{1-R} \rceil$ and
$C_{i}=\exp(-\lambda^{i}N^{i}_{T}(n))$.

\begin{figure}[htbp]
\begin{center}
\includegraphics[width=10cm]{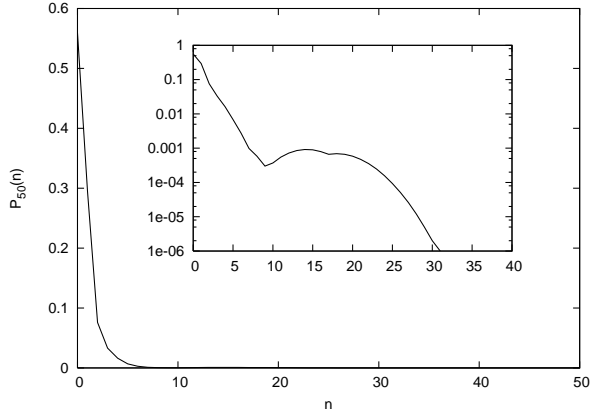}
\caption{\label{fig:Dist_8_30}
Plot of implied default distribution for fifty Japanese companies
on August 30, 2005.}
\end{center}
\end{figure}

The six Lagrange multiplier were calibrated so that the
condition $<N_{T}^{i}>=N_{T,Implied}^{i}$ is satisfied.
We use the simulated annealing method and fix these parameters.
Figure \ref{fig:Dist_8_30} shows the result of fitting 
eq. (\ref{eq:Dist}) to iTraxx-CJ data on August 30, 2005.
About the convergence, it is satisfactory and 
all premiums
are recovered within $1\%$. From the inset figure, which shows a
semilog plot of the distribution, we see a hunchy structure
or a second peak.
$P_{N}(n)$ decreases monotonically
up to the fourth tranche ($n \le 9$), then  $P_{N}(n)$
 begins to increase. In the fifth tranche $n_{H}^{4}=10 < n \le 
n_{H}^{5}=17$, $P_{N}(n)$  has a peak and then decreases to zero. 
We also see some joints between tranches at $n_{H}^{j}$.
The latter is an artifact of the maximum entropy principle.

\section{Comparison with Popular Probabilistic Models}
\label{comparison}

In this section, we compare the behaviors of the loss probability
function $P_{N}(n)$ of some popular probabilistic models with 
the implied loss distribution function 
from the
viewpoint of the correlation structure. In particular, we 
focus on $\rho_{i,0}$ and $p_{i,0}$.
As probabilistic models, we consider the next three models.
These models are defined by the mixing function $f(p)$ that express
the joint probability function $X_{i,j}$ as \cite{Schonbucher,Hisakado}
\begin{equation}
X_{i,j}=\int f(p'){p'}^{i}(1-p')^{j} dp'.
\end{equation}
We choose the Gaussian copula (GC) model, which is a standard model in
financial engineering \cite{Hull4}, 
the beta binomial distribution (BBD), which is the benchmark model among 
exchangeable correlated binomial models \cite{Hisakado} and the long-range 
Ising (LRI) model \cite{Kitsukawa}. 
The reason for adopting LRI, instead
of the Ising model on some lattice, is that 
in financial engineering
all obligors are usually assumed to be related to each other 
with the same strength and that the network structure is uniform.
In addition, the long-range Ising model can be expressed as a
superposition of two binomial distributions for 
sufficiently large $N$ and it is very tractable.

\begin{enumerate}
\item Gaussian Copula (GC) Model.

The model incorporates the default correlation $\rho_{d}$ by a
common random factor $Y$ and an asset correlation $\rho_{a}$. 
If the factor $Y$ is fixed as $Y=y$, the variables $X_{i}$ become
independent with the probability $\mbox{Prob.}(X_{i}=1)=p(y)$.
The explicit form of the mixing function is
\begin{equation}
f(p(y))=\Phi(\frac{K-\sqrt{\rho_{a}}y}{\sqrt{1-\rho_{a}}}).
\end{equation}
Here, $K=\Phi^{-1}(p_{d})$ with the 
normal cumulative function $\Phi(K)$
and $Y$ obeys the normal distribution $Y\sim N(0,1^{2})$.
$X_{i,j}$ are then given as
\begin{equation}
X_{i,j}=<p(y)^{i}(1-p(y))^{j}>_{Y}.
\end{equation}
$<\hspace*{0.5cm}>_{Y}$ denotes the expectation value over the random
variable $Y$. In order to estimate $\rho_{d}$, 
we use the relation 
$\rho_{d}=\frac{X_{2,0}-p_{d}^{2}}{p_{d}(1-p_{d})}$.

\item Beta Binomial Distribution (BBD) Model.

The mixing function $f(p')$ is the beta distribution.
\begin{equation}
f(p')=\frac{{p'}^{\alpha-1}(1-p')^{\beta-1}}{\mbox{B}(\alpha,\beta)}.
\end{equation}
Here $\mbox{B}(\alpha,\beta)$ is the beta function.
$X_{i,j}$ are given as
\begin{equation}
X_{i,j}=\frac{\mbox{B}(\alpha+i,\beta+j)}{\mbox{B}(\alpha,\beta)}.
\end{equation}
It is easy to show that  $p_{0,0}=p_{d}=\frac{\alpha}{\alpha+\beta}$
and $\rho_{0,0}=\rho_{d}=\frac{1}{\alpha+\beta+1}$.

We note that BBD is the benchmark model among exchangeable correlated
 binomial models. 
$\rho_{i,j}$ depend on $i,j$ through the form  $i+j$ as 
$\rho_{i,j}=\frac{\rho_{d}}{1+(i+j)\rho_{d}}$.
 As the result, 
$p_{i,j}$ becomes a linear  function of $i$ for the fixed $i+j=k$ as
\[
p_{i,j}=\frac{p_{d}(1-\rho_{d})+i\cdot \rho_{d}}{1+(k-1)\rho_{d}}. 
\]
BBD is the ``linear'' model \cite{Hisakado} that is why we call it
the benchmark model.
One can see the nonlinearity of other models 
by checking the differences of $\rho_{i,j}$ and $p_{i,j}$ 
from those of BBD.

\item Long-Range Ising (LRI) Model \cite{Mori}.

The mixing function $f(p')$ is the superposition of two $\delta$
functions $\delta(p'-p)$ and $\delta(p'-(1-p))$.
\begin{equation}
f(p')=(1-\alpha)\cdot \delta(p'-p)+\alpha\delta(p'+(1-p)).
\end{equation}
$X_{i,j}$ are given as
\begin{equation}
X_{i,j}=(1-\alpha)p^{i}(1-p)^{j}+\alpha (1-p)^{i}p^{j}.
\end{equation}
It is easy to show that 
$p_{0,0}=p_{d}=(1-\alpha)p+\alpha(1-p)$
and $\rho_{0,0}=\rho_{d}=\frac{\alpha(1-\alpha)(2p-1)^{2}}{p_{d}(1-p_{d})}$.
\end{enumerate}

\begin{figure}
\begin{center}
\includegraphics[width=10.0cm]{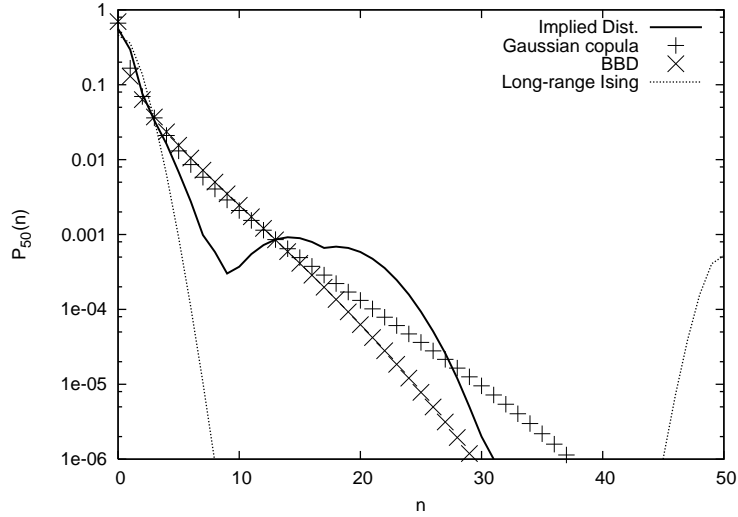}
\caption{\label{fig:Dist_comp} 
Plot of implied default distribution (Fig.\ref{fig:Dist_8_30}) 
and the probability functions of popular 
pricing models. Implied distribution (solid), GC
 ($\times$), BBD ($+$) and LRI (dashed-dotted line).
We set $N=50$,$p=1.65\%$ and $\rho_{d}=6.55\%$.
}
\end{center}
\end{figure}

Figure \ref{fig:Dist_comp} shows a plot of the implied distribution of the 
previous section 
with the probability function $P_{N}(n)$ of the above  
three models.
The models have three parameters : the number of variables $N$,
the default probability 
$p_{d}$ and default correlation $\rho_{d}$. We set them with
the same values of the implied distribution as $N=50$,
$p_{d}=1.65 \%$ and $\rho_{d}=6.55\%$. 
We see that all three models give poor fits to the implied distribution. 
GC and BBD 
show  a monotonic dependence on $n$. The LRI model has a nonmonotonic
dependence and has a hump at $n=N$.

\begin{figure}
\begin{center}
\includegraphics[width=10cm]{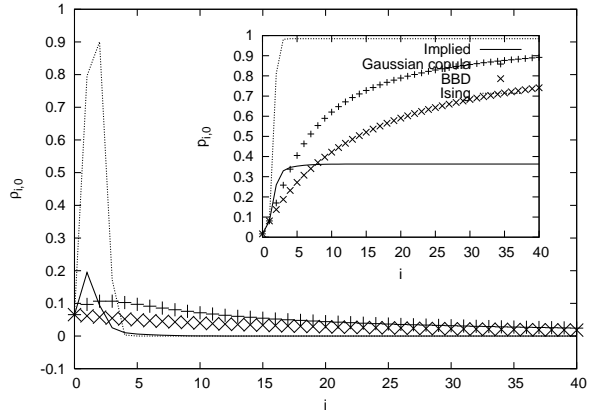}
\caption{\label{fig:Rho_i0_P_i0} Plots of $\rho_{i,0}$
and $p_{i,0}$ (inset figure).}
\end{center}
\end{figure}

Next, we compare their correlation structures.  
Figure \ref{fig:Rho_i0_P_i0} depicts $\rho_{i,0}$ and $p_{i,0}$. 
$\rho_{i,0}$ for GC has a 
low peak and decays to zero slowly. BBD's $\rho_{i,0}$ decays
slowly as $\rho_{i,0}=\frac{\rho_{d}}{1+i \rho_{d}}$.  
On the other hand,
LRI's $\rho_{i,0}$ rapidly increases
to $1$ and decays rapidly to zero.
This behavior means that GC is weakly nonlinear and 
LRI is strongly nonlinear.

As for $p_{i,0}$, recall the relation 
$p_{i+1,0}=p_{i,0}+(1-p_{i,0})\rho_{i,0}$ (eq.(
\ref{consist_p})).
With the same $p_{d}$ and $\rho_{d}$,
we have $p_{0,0}=p_{d}$ and $p_{1,0}=p_{d}+(1-p_{d})\rho_{d}$. All
curves $(i,p_{i,0})$ 
go through the two points $(0,p_{d})$  and $(1,p_{d}+(1-p_{d})\rho_{d})$. 
As $\rho_{i,0}$ for $i\ge 1$ differs among the models and 
the implied one, the curves $(i,p_{i,0})$
depart from each other for $i\ge 2$. 
LRI's $\rho_{i,0}$ rapidly increases
to $1$. $p_{i,0}$ also increases to 1 rapidly. For $i=3$, 
$p_{i,0}\simeq
1$ and this means that all the obligors always default simultaneously
if three of them 
are defaulted, which is the  biggest avalanche.
As the result, $P_{N}(n)$ has a hump in its tail $n=N$.
GC's $p_{i,0}$ and BBD's $p_{i,0}$
 increase to 1 with $i$ slowly. The distribution of 
the size of avalanches should be very wide and $P_{N}(n)$ comes
to have a long tail.

The implied one's $\rho_{i,0}$ has a medium peak at $i=1$ and then
rapidly decreases to zero for $i\ge 5$. Comparing it with those of BBD,
we see that the implied loss distribution function is nonlinear.
Its behavior is completely different from those of both GC and BBD. 
$p_{i,0}$ increases rapidly with $i$ compared with
 GC and BBD and  soon saturates to
 some maximum value $\simeq 0.35$ at $i=5$. 
The credit market expects that if more than 5 defaults 
occur, the obligors default almost independently.  
 The size of an avalanche of simultaneous defaults is smaller
than that of the Ising model. However, the probability that a
medium-size of avalanche of defaults occurs is large compared with the 
GC and BBD.

\begin{figure}
\begin{center}
\includegraphics[width=10cm]{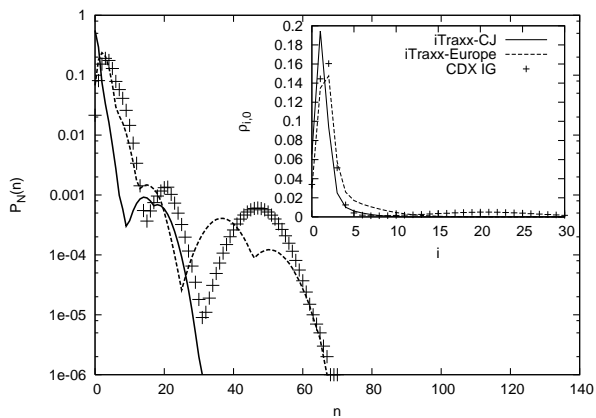}
\caption{\label{fig:Dist_Corr_JEU} 
Implied default distribution on August 30, 2005. iTraxx-CJ (in solid),
iTraxx-Europe (dashed) and CDX IG (+).}
\end{center}
\end{figure}

We also studied the correlation structures of the implied loss 
functions  of iTraxx-Europe and CDX IG (U.S.A.), which are 
CDOs of European and American companies ($N=125$) \cite{Hull4}. 
The implied distributions and  $\rho_{i,0}$ are plotted in 
Fig. \ref{fig:Dist_Corr_JEU}. The implied loss functions  are more
complex than that of iTraxx-CJ. $\rho_{i,0}$ shows the same singular 
behavior with those of iTraxx-CJ and the singular behavior 
seems to be a  universal 
property.

About the origin of the singular behavior of $\rho_{i,0}$,
we point out two possibilities. The first is that the probabilistic
rule that governs the defaults of obligors is essentially new 
and nonlinear. The second is that the nonuniform network
 structure of the dependency relation of the obligors is reflected 
in $\rho_{i,0}$.
If the network structure is not uniform, it affects
 the resulting correlation structure. As a result, $\rho_{i,0}$ 
looks singular compared with those of the models on the uniform network.

\begin{table}
\caption{\label{tab:tranche}Implied 
tranche correlations and  entropy maximum correlation for 5-year
 iTraxx-CJ tranches on August 30, 2005.
}
\begin{center}
\begin{tabular}{cccccc}
$\{0\%,3\%\}$&$\{3\%,6\%\}$&$\{6\%,9\%\}$&$\{9\%,12\%\}$&$\{12\%,22\%\}$&Entropy\\
\hline
13.5\% &1.20\% &2.58\% &4.95 \% &9.71 \% &6.55 \% \\
\end{tabular}
\end{center}
\end{table}

At last we comment on the tranche (compound) 
correlation, which is the  standard correlation measure 
in financial engineering \cite{Finger}.
The method suggests the correlation $\rho_{d}^{i}$ so that 
the expected loss equals the expected payoffs for the $i-$th tranche, 
 it is called ``tranche correlations''.
The expected values are estimated with GC. 
Table \ref{tab:tranche} shows the 
 tranche correlations for the quotes of iTraxx-CJ on August 30, 2005. 
In the last column, we show the maximum entropy value derived from 
the implied default distribution. 
As we showed previously, the GC
gives a poor fit to the implied distribution. The tranche
correlations are completely different from the entropy maximum value.
In addition, it depends on which tranche the correlation is estimated.
Such a dependence is known as a ``correlation smile'' \cite{Andersen}. 
We think that the ``true'' default correlation is approximately given by 
the maximum entropy value and that tranche correlations are an artifact 
of using GC to fit the market quotes. 
As long as the probabilistic model gives a poor fit to the market quotes,
the default correlation varies among the tranches. 
This is the origin of the ``correlation smile''.

\section{Conclusions}
\label{conclusion}

We show how to 
estimate the conditional
probabilities $p_{i,j}$ and correlations $\rho_{i,j}$
from $P_{N}(n)$. 
BBD is the benchmark model among exchangeable 
correlated binomial models and $\rho_{i,j}$ behave as 
$\rho_{i,j}=\frac{\rho_{d}}{1+(i+j)\rho_{d}}$. 
If the obtained $\rho_{i,j}$ depends on $i,j$, which is 
considerably different
from those of BBD, there are two possibilities.
The first one is that the probabilistic rule that governs 
$X_{i}$s
is strongly nonlinear.
The second one is that the assumption of the exchangeablity  is wrong.
The network structure of the dependency relation among $X_{i}$s 
is nonuniform. 

We have inferred the loss probability function for multiple defaults
based on the market quotes of CDOs and the maximum entropy principle.
The profile is completely different from those of some popular 
probabilistic models, namely GC, BBD and LRI. 
$\rho_{i,0}$ has a medium peak and then rapidly decreases 
to zero for $i\ge 5$.
The origin of the singular behavior
can be attributed to the above two possibilities. 

In order to clarify the mechanism of the singular behavior
of $\rho_{i,0}$, it is necessary to study correlated binomial 
models on networks. In particular, the dependence of $\rho_{i,0}$
on the network structure should be understood.
 Recently, the authors 
have shown how to construct a linear correlated binomial model
on networks in general \cite{Mori2}. By applying the method of the present
paper to the model, it is possible to understand the relation between 
the network structure and the correlation structures.
More detailed studies of real companies' dependency structures 
have been performed recently \cite{Fujiwara}. 
Instead of the implied loss function, a real
loss distribution function has been estimated. 
Promoting these studies, we think that it is possible 
to understand the dependency structure of multiple defaults and 
to propose a theoretical model of the pricing of CDOs.

\appendix

\section{Probability Function and Correlation Structure}
\label{structure}

\begin{figure}[htbp]
\begin{center}
\includegraphics[width=9cm]{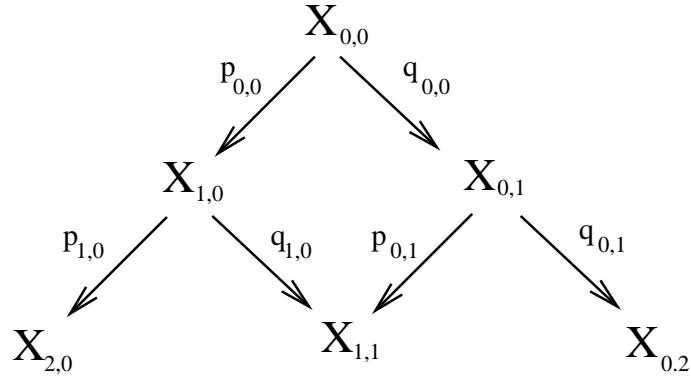}
\end{center}
\caption{Pascal's triangle representation 
of $X_{i,j}$ up to $i+j\ \le 2$ and $p_{i,j},q_{i,j}$ .
$X_{0,0}=<1>$, $X_{1,0}=<X_{1}>=p$, $X_{0,1}=<1-X_{1}>=1-p=q$ etc.}
\label{pascal_old}
\end{figure}

In this section, we explain the relation between $P_{N}(n)$ and 
the correlation structures $p_{i,j}$ and $\rho_{i,j}$.
We introduce the products
of $X_{i}$ and $1-X_{j}$, which exhaust all observables of the system.
\begin{equation}
\Pi_{i,j}=\prod_{i'=1}^{i}X_{i'}\prod_{j'=i+1}^{i+j}(1-X_{j'})
\end{equation}
The following definitions are their unconditional and conditional 
expectation values (see Fig. \ref{pascal_old}.).
\begin{eqnarray}
X_{i,j}&=&<\Pi_{i,j}>. \\
p_{i,j}&=&<X_{i+j+1}|\Pi_{i,j}=1>=<\Pi_{i+1,j}|\Pi_{i,j}>=
\frac{X_{i+1,j}}{X_{i,j}}.
\\
q_{i,j}&=&<1-X_{i+j+1}|\Pi_{i,j}=1>=<\Pi_{i,j+1}|\Pi_{i,j}>=
\frac{X_{i,j+1}}{X_{i,j}}.
\end{eqnarray}
Here $<A|B>$ means the expectation value of the random variable
$A$ under the condition that $B$ is satisfied. 
$X_{0,0}=1$,
$X_{1,0}=p_{0,0}$ and $X_{0,1}=1-p_{0,0}=q_{0,0}$. 
All information of the model is contained in $X_{i,j}$. 
The joint probability $P(x_{1},x_{2},\cdots,x_{N})$ with 
$\sum_{i'=1}^{N}x_{i'}=n$ is given by $X_{n,N-n}$.
The probability function $P_{N}(n)$ is given as
\[
P_{N}(n)\equiv \mbox{Prob}(\sum_{i=1}^{N}X_{i}=n)=
{}_{N}C_{n} \cdot X_{n,N-n}. 
\] 

We also introduce the conditional correlation
\begin{equation}
\mbox{Corr}(X_{i+j+1},X_{i+j+2}|\Pi_{i,j}=1)=\rho_{i,j}.
\end{equation}
The correlation between $X_{i}$ and $X_{j}$
is defined as 
\begin{equation}
\mbox{Corr}(X_{i},X_{j}) 
\equiv \frac{<X_{i}X_{j}>-<X_{i}><X_{j}>}{\sqrt{<X_{i}>(1-<X_{i}>)<X_{j}>
(1-<X_{j}>)}} . \label{pearson} 
\end{equation}
Its conditional ones are  defined by replacing
expectation values with conditional expectation values.

The conditional quantities $p_{i,j},q_{i,j}$ and $\rho_{i,j}$ 
must obey the recursive relations from
eqs. (\ref{consist_p})-(\ref{consist_rho}). 
The reason is that the following two relations must hold  for the system
to  be consistent. 
The first one is $p_{i,j}+q_{i,j}=1$ for any $i,j$, because of the identity
$<1|\Pi_{i,j}=1>=<X_{i+j+1}+(1-X_{i+j+1})|\Pi_{i,j}=1>=1$. 
The second one is the commutation relation
\begin{equation}
q_{i+1,j}\cdot p_{i,j}=p_{i,j+1}\cdot q_{i,j}=\frac{X_{i+1,j+1}}{X_{i,j}}.
 \label{consist}
\end{equation}
These two relations are guaranteed to hold when $p_{i,j},q_{i,j}$ and 
$\rho_{i,j}$ satisfy the above consistency relations.

We explain the meaning of these quantities.
The first one is $p_{0,0}$, the unconditional expectation value of
$X_{i}$. Its meaning is clear and it is the probability that $X_{i}$
takes 1. In the context of a credit portfolio problem, it is the default
probability $p_{d}$. It is easy to estimate it from $P_{N}(n)$ as
\begin{equation}
p_{d}=p_{0,0}=<X_{i}>=<n>/N. 
\end{equation}
The unconditional correlation $\rho_{0,0}$ is the 
default correlation $\rho_d$ in the credit risk context. 
It is also easy to estimate it as
\begin{equation}
\rho_{d}=\rho_{0,0}=\frac{<n^{2}-n>/N(N-1)-p_{d}^{2}}{p_{d}(1-p_{d})}. 
\end{equation}
Its estimation is important in the evaluation of the prices of 
credit derivatives. One reason is that it is related to the conditional
default probability $p_{1,0}$ from eq. (\ref{consist_p}) as
\begin{equation}
p_{1,0}=p_{d}+(1-p_{d})\rho_{d}.
\end{equation}
If one obligor is defaulted, the default probability $p_{d}$
changes to $p_{1,0}$. 
The second reason is that it gives the
simultaneous default probability for $X_{i}$ and $X_{j}$ as
\begin{equation}
\mbox{Prob.}(X_{i}=1,X_{j}=1)=p_{d}^{2}+p_{d}(1-p_{d})\rho_{d}.
\end{equation}
Usually,  $p_{d}$ is small and  the simultaneous default probability
is mainly governed by the second term.

\begin{figure}[htbp]
\begin{center}
\includegraphics[width=9.0cm]{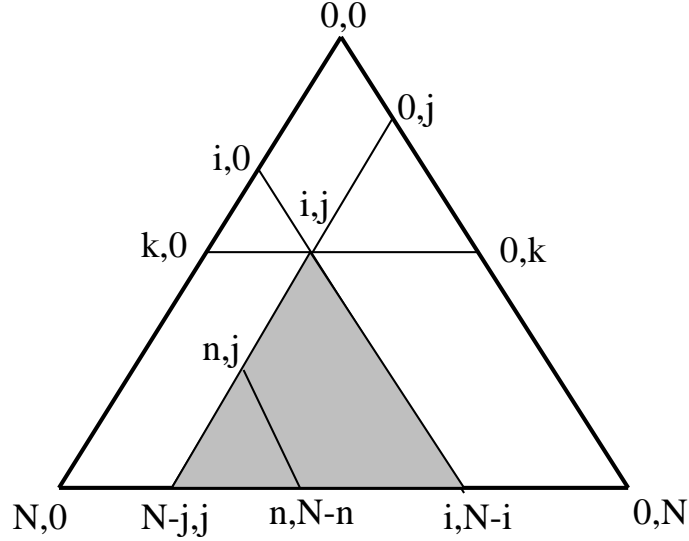}
\caption{\label{fig:Pascal_3}
Restricted Pascal Triangle with the vertex $(i,j),(N-j,j)$ and
 $(i,N-i)$. $p_{l,m}$ and $\rho_{l,m}$ in the triangle 
govern the behavior of the system under the condition $\Pi_{i,j}=1$.

}
\end{center}
\end{figure}

Regarding $p_{i,j}$ with $i\hspace*{0.1cm} \mbox{or}
\hspace*{0.1cm}  j>0$, we note one point. 
From the definition, $p_{i,j}$ means the default probability under the 
condition $\Pi_{i,j}=1$. $\rho_{i,j}$ also means the default correlation  
in  the same 
situation. $p_{l,m}$ with $l \ge i$ and $m\ge j$ are closely related to 
the default probability
function $P_{N-(i+j)}(n-(i+j)|\Pi_{i,j}=1)$. We write $k=i+j$ and the next
relation holds for $n \ge k$.
\begin{equation}
P_{N-k}(n-k|\Pi_{i,j}=1)={}_{N-k}C_{n-k}\cdot 
<\prod_{l=1}^{n-k}X_{k+l}\prod_{m=1}^{N-n}(1-X_{n+m})
|\Pi_{i,j}=1>
\end{equation}
We evaluate the expectation value with $p_{l,m}$ and $q_{l,m}$, and  we get
\begin{equation}
P_{N-k}(n-k|\Pi_{i,j}=1)={}_{N-k}C_{n-k}\cdot
\prod_{l=0}^{n-k-1}p_{i+l,j}\prod_{m=0}^{N-n-1}q_{n-k+i,j+m} 
\hspace*{0.3cm}n\ge k.
\end{equation}
This relation indicates that the Pascal Triangle with the vertex 
$(i,j),(N-j,j)$, and $(i,N-i)$ contains all information for 
the case $\Pi_{i,j}=1$ (See Fig. \ref{fig:Pascal_3}). In order to know the 
loss probability function under the condition $\Pi_{i,j}=1$, 
we only need to know $p_{l,m}$ and $q_{l,m}$ 
in the restricted Pascal Triangle.

The $i$-dependence of $\rho_{i,0}$ and $p_{i,0}$ is 
closely related to the 
behavior of the probability function $P_{N}(n)$ for $n\ge i$.
By the relation, we can understand the cascading structure of 
the simultaneous defaults. 
Hereafter, as we are interested in the
credit risk problem, we assume that $p_{0,0}=p_{d}$ is small.

First, we note that $P_{N}(n)$ can be expressed in the following form
 for $n\ge i$.
\begin{equation}
P_{N}(n)=\frac{{}_{N}C_{n}}{{}_{N-i}C_{n-i}}\cdot X_{i,0} \cdot
P_{N-i}(n-i|\Pi_{i,0}=1).  \label{eq:partial}
\end{equation}
The derivation is based on the following relation.
\begin{eqnarray}
&&P_{N-i}(n-i|\Pi_{i,0}=1)
={}_{N-i}C_{n-i}\times <\prod_{l=i+1}^{n}X_{l}\prod_{m=n+1}^{N}
(1-X_{m})|\Pi_{i,0}=1> \nonumber \\
&=&{}_{N-i}C_{n-i}\times <\prod_{l=1}^{n}X_{l}\prod_{m=n+1}^{N}
(1-X_{m})|\Pi_{i,0}=1> \nonumber \\
&=&{}_{N-i}C_{n-i}\times <\prod_{l=1}^{n}X_{l}\prod_{m=n+1}^{N}
(1-X_{m})>/<\prod_{l=1}^{i}X_{i}> \nonumber \\
&=&{}_{N-i}C_{n-i}
\times \frac{P_{N}(n)}{{}_{N}C_{n}}/X_{i,0}.
\end{eqnarray}
Equation (\ref{eq:partial}) tells us about 
the behavior of $P_{N}(n)$ for $n \ge i$. 
 
We classify the behavior $\rho_{i,0}$ into two cases.
\begin{enumerate}
\item Short-tail case :

The probability function $P_{N}(n)$ develops a short tail
 in the case where $\rho_{i,0}$ rapidly decreases with $i$ and 
$\rho_{i,j}=0$ for $i\ge k$ and $j\le i-k$ with $k<<N$.
For the case  $\Pi_{i,j}=0$ with $i\ge k$ and $j\le i-k$, 
all variables become independent.
$X_{i,j}$ is estimated as
\[
X_{i,j}=
X_{k,0}\cdot <\Pi_{i,j}|\Pi_{k,0}=0>
=X_{k,0}\cdot p_{k,0}^{i-k}\cdots q_{k,0}^{j}.  
\]
The probability function $P_{N}(n)$ for $n\ge k$ becomes 
\[
P_{N}(n)={}_{N}C_{n}\cdot \frac{X_{k,0}}{p_{k,0}^{k}}\cdot p_{k,0}^{n}
\cdot q_{k,0}^{N-n}. 
\]
$P_{N}(n)$ becomes proportional to 
the  binomial distribution $\mbox{Bi}(N,p_{k,0})$ and 
has a short tail. It has a hump at $n\simeq N\cdot p_{k,0}$.
In particular, if $p_{k,0}\simeq 1$, the 
probability function has a hump at $n=N$.

\item Long-tail case : 

The probability function has a long tail in the case where
$\rho_{i,0}$ is  small and gradually 
decreases with $i$. The random variables are weakly coupled.
The $i$-dependence of 
$p_{i,0}$ is given by $p_{i+1,0}=p_{i,0}+(1-p_{i,0})\rho_{i,0}$ and 
$p_{i,0}$ gradually increases with $i$. If we assume $\rho_{i,0}=0$, 
$P_{N}(n)$ becomes proportional to the binomial distribution 
$\mbox{Bi}(N,p_{i,0})$ for $n\ge i$.
\[
P_{N}(n)={}_{N}C_{n}\cdot \frac{X_{i,0}}{p_{i,0}^{i}}\cdot p_{i,0}^{n}
\cdot q_{i,0}^{N-n}  \hspace*{0.3cm}\mbox{for}
\hspace*{0.3cm} n\ge i.
\]
However, $\rho_{i,0}$ is not zero
and $p_{i,0}$ gradually increases with $i$. 
For $n\ge i+1$, $P_{N}(n)$ behaves as
\[
P_{N}(n)={}_{N}C_{n}\cdot 
\frac{X_{i+1,0}}{p_{i+1,0}^{i}}
\cdot 
p_{i+1,0}^{n}
\cdot q_{i,0}^{N-n}  \hspace*{0.3cm}\mbox{for}
\hspace*{0.3cm} n\ge i.
\]
$X_{i+1,0}=X_{i,0}\cdot p_{i,0}$, we have 
\[
P_{N}(n)={}_{N}C_{n}\cdot 
\frac{X_{i,0}}{p_{i+1,0}^{i}}
\cdot 
\frac{p_{i,0}}{p_{i+1,0}}
\cdot 
p_{i+1,0}^{n}
\cdot q_{i,0}^{N-n}  \hspace*{0.3cm}\mbox{for}
\hspace*{0.3cm} n\ge i.
\]
As $p_{i+1,0}>p_{i,0}$, the overall scale $\frac{X_{i,0}}{p_{i+1,0}^{i}}
\cdot 
\frac{p_{i,0}}{p_{i+1,0}}
$ is smaller than $\frac{X_{i,0}}{p_{i,0}^{i}}$. 
Apart from the overall factor,
$P_{N}(n)$ becomes proportional 
to $\mbox{Bi}(N,p_{i,0})$ with a larger $p_{i,0}$ for a larger $i$.
Compared with that in the 
short-tail  case, the decrease in $P_{N}(n)$ with $n$
is milder and $P_{N}(n)$ has a longer  tail.

\end{enumerate}

\end{document}